\documentstyle[psfig]{mn}

\begin{document}

\title[Double Starbursts: Mergers]{Double Starbursts Triggered by Mergers in Hierarchical Clustering Scenarios}

\author[Tissera et al.]
{P.B. Tissera$^1$, R. Dom\'{\i}nguez-Tenreiro$^2$, C. Scannapieco$^{1}$
  and A. S\'aiz$^2$\\
  $^1$I.A.F.E., Casilla de Correos  67, Suc.\ 28, Buenos
  Aires, 1428, Argentina\\
  $^2$Departamento de F\'{\i}sica Te\'orica, C-XI. Universidad
  Aut\'onoma de Madrid, Madrid, E-28049, Spain\\
 }

\maketitle

\begin{abstract}
We  use cosmological SPH simulations to  study  the effects of mergers in 
the star formation history of galactic objects in 
hierarchical clustering scenarios.  We find that during some merger events, gaseous discs  can experience
two starbursts:
the first one during the orbital decay phase, due to gas inflows driven as the satellite approaches,
and the second one, when the two baryonic clumps collide. A trend
for these first induced starbursts to be more efficient at transforming
the gas into stars is also found. We detect that systems which
do not experience early gas inflows have 
 well-formed stellar bulges and 
more concentrated  potential wells, which seem to be  responsible for preventing further 
gas inward  transport triggered by tidal forces.
The potential wells concentrate due to the accumulation of baryons in the
central regions and of  dark matter 
 as the result of the pulling in by baryons.
The   coupled evolution of the dark matter and
baryons would lead to an evolutionary sequence during which 
systems with shallower total  potential wells suffer early gas inflows during
the orbital decay phase that help to feed their  central  mass concentration,
 pulling in dark matter and  contributing to
build up  more stable systems. 
Within this scenario, starbursts  triggered
by early gas inflows are more likely to occur at early stages of
evolution of the systems
and to be an important contributor to the formation of stellar bulges.
Our results
constitute the first proof that bulges can form as the product
of collapse, collisions and secular evolution in a
cosmological framework, and they
are consistent with a rejuvenation of the stellar population
in bulges at intermediate $z$ with, at least, $50 \%$ of the stars
(in SCDM) being formed at high $z$.

\end{abstract}

\begin{keywords}
galaxies: evolution - galaxies: formation - galaxies: 
interactions  - cosmology: theory -
cosmology: dark matter
\end{keywords}

\section{Introduction}

The physical mechanisms that  trigger and regulate
star formation in galaxies are thought to be of different
nature and act at different scales (e.g., Kennicutt 1998). Among 
them, mergers and interactions are known to play a crucial
role as it has been demonstrated from observations of
high emission levels of
 $H_{\alpha}$ (e.g., Kennicutt et al.\ 1987), far infrared (e.g., Solomon \& Sage 1988; Sanders \& 
Mirabel 1996) and radio
continuum (e.g., Condon et al.\ 1982)  in interacting systems.
Recent observations of nearby galaxy  pairs by Barton, Geller \& Kenyon
(2000; see also Donzelli \& Pastoriza 1997) found that interactions
during close encounters can  be statistically 
correlated with  enhancements of the  
star formation activity  
lasting more than $10^8$ yr.
It has also been  observed that moderately luminous starbursts
in the nearby Universe may occur in disc galaxies that show
only mild external perturbations  
originated in moderate interactions (Gallagher et al.\ 2001).
Regarding our Galaxy, a new study of the chemical composition 
of the solar neighbourhood by 
Rocha-Pinto, Maciel \& Flynn (2000)
supports the possible existence
of three main starburst episodes,   lasting each one few Gyr,
which can be associated with nearby passage of the Milky Way
galaxy satellites. 
Remarkably, two
of such starbursts show evidences of being 
associated to only one of those close encounters.
Recent 
observations of high redshift ($z$) objects
show an increase of  morphologically disturbed and interacting objects 
with $z$, 
accompanied by an increase of the star formation 
activity (e.g., Bouwer, Broadhurst \& Silk 1998; Driver et al.\ 1998; Brinchmann et al.\ 1998; Sawicki \& Yee 1998; Le F\`evre et al.\ 2000). 
Although it is not yet clear at what extent these data are fully
consistent with predictions of models based on hierarchical clustering,
 all of them agree in supporting a  scenario  in which interactions/mergers
and star formation increase with $z$
 (Flores et al.\ 1999; Steidel et al.\ 1999).

Understanding the nature of galaxies as a function of $z$ and the interface between galaxy formation and structure formation,
requires to be able to describe how star formation proceeds in galactic objects at different stages of
evolution, and how this activity is affected by the formation of the structure
at larger scales.
This is a complex task taking into account that 
as a galaxy evolves, it may react to the same
physical mechanism (i.e., such as mergers/interactions) 
in a different way, depending on its
astrophysical and dynamical  properties (i.e., formation of a bulge, baryonic/dark matter dominated potential well
at the centre, availability of gas, etc), and, consequently, on its previous evolutionary history.

In a hierarchical clustering scenario, galactic objects
form
 by the aggregation of substructure. Assuming this
model as the more feasible one involves the understanding of the role played by mergers and interactions as
an object evolves.
In particular, pre-prepared simulations of merging
galaxies show that, during violent events,
strong gas inflows can be triggered, feeding star formation
activity at the centre (e.g., Noguchi \& Ishibashi 1986; Noguchi 1991; Mihos, Bothum \& 
Richstone 1993; Mihos 1992; Barnes \& Hernquist 1991, 1996). Furthermore,
 Mihos \& 
Hernquist (1994, 1996 hereafter MH96) demonstrated 
that during merger events,
a bulge-less disc galaxy is 
more susceptible to growing bar instabilities that 
might trigger  gas inflows towards its centre, even before the
actual collision of the baryonic clumps occurs, than a 
bulge-disc system.
In the first  case, a double star formation burst was identified, with a
first component triggered by  the strong
tidal fields developed during the orbital decay phase (hereafter  ODP), 
and a second one triggered when  the baryonic cores collided.
 The relative strength
of the double bursts was shown to depend on the compactness of the bulge and
on the availability of gas at the time of the merger.
Conversely, a disc galaxy with a stellar bulge 
was found to be  more stable, and 
only when the actual collision of the
baryonic cores occurred, a starburst was produced. These authors
directly link the response of the systems during a merger to the 
characteristics
of their potential wells.

However, in these pre-prepared mergers, the great variety of phenomena
involved in the complex processes of galaxy-like objects (GLOs) building-up
from the field of primordial fluctuations at high $z$
up to $z=0$, cannot be followed. Such phenomena
include collapse events, mergers and interactions with close
neighbours, accretion of smaller objects, gas infall at
scales of some hundreds of kpcs (from the environment to the halo or from
the halo to the disc),
gas inflow at scales of tens of kpcs (within discs and from
discs to the bulge-like central concentrations), etc.
All of them play a role in the assembly of galaxy-like objects,
and, moreover, they interact among themselves in a non-trivial way.
Pre-prepared mergers are very detailed experiments, but
which only describe a particular phase of galaxy formation
from initial conditions that are generally set by hand.
Moreover, mergers in these experiments are
taken {\it in isolation\/}, and not in connection with the
other processes involved in galaxy formation.
On the contrary, with a self-consistent approach (i.e.,
following the assembly of galaxy-like objects from the
field of primordial fluctuations
at high $z$ up to $z=0$,
in the framework of a cosmological
model), mergers can be studied in connection with
other mechanisms, that play a fundamental role in the merger
process itself, such as gas infall at 
some hundreds of kpc scales
and gas inflows in discs. Also,
one can analyze how  the initial conditions that are
usually considered in pre-prepared simulations
(at scales of about their virial radii, i.e., some hundreds of kpcs),
appear, along the evolution  of the objects, 
in connection with the evolution of their environment at larger scales.
 At these larger scales, evolution is
determined  by
the global evolution of the cosmological model. Of particular
interest in this regard is to learn how bulges are assembled,
and how the growth of a given bulge at a given $z$ might prevent
it from growing
at lower $z$s.

By using cosmological hydrodynamical
simulations, Dom\'{\i}nguez-Tenreiro, Tissera \& 
S\'aiz (1998, hereafter DTTS98) and 
 Tissera, S\'aiz \& Dom\'{\i}nguez-Tenreiro (in preparation)
 analyzed how  gas inflows can
be generated during the evolution of a 
typical galaxy-like object in a hierarchical clustering
model, as it forms by the  merger
of substructure. 
These authors showed that the gas with non-zero angular
momentum 
 always  tends  to form a disc, but
the survival probability of these discs depend
on their stability properties.  Their findings confirmed that gas inflows
 can be easily triggered during violent events
if the axi-symmetrical character of the potential
well is not assured by 
a  central mass concentration, as for example,
 a dark matter halo
dynamically dominating at the centre, or   
by  the presence  of a compact stellar bulge
  (e.g.,Toomre \& Toomre 1972; Martinet 1995 and
references therein). 
These gas inflows  can be mild or catastrophic, depending
on the internal properties of the disc-like systems. 
Catastrophic inflows can destroy the discs and are
 easily induced in 
the absence of a compact central mass concentration
 (DTTS98; Steinmetz \& Navarro 1999).
However, if  stellar bulges are allowed to form,  discs
with observational counterparts can be formed in
hierarchical clustering scenarios (S\'aiz et al.\ 2001, hereafter Sal01).

Tissera (2000a, hereafter Tis00) studied the star 
formation history of galactic-like objects
 in relation to their
merger histories within a cosmological context. This author   found
that minor and major mergers may trigger starbursts as powerful,
 regardless of their relative masses, if there
is enough available gas  in the system. Only the starbursts
efficiencies, defined as the fraction of the available
gaseous mass  in the system that is actually transformed into stars, were found to depend on the relative masses of
the colliding objects.
The stellar masses formed and starbursts (SBs) durations measured in the simulations
are in agreement with observational results (e.g., Barton et al.\ 2000;
Glazebrook et al.\ 1998).
Even more, Tis00 estimated a change  of up to 0.6 magnitudes in the
colours of the simulated galaxies
 due to
mergers, 
 consistent
with observational results reported by Le F\`evre et al.\ (2000).
It was also pointed out by the author that  during certain merger
events,  double SBs were present. 
If these phenomena were a common physical process, then, understanding how
they could be induced in a hierarchical clustering scenario 
 is  very
relevant to the study of galaxy formation.
 The physical interpretation of several observations
of galaxy properties, such as luminosity function or spatial
distributions,  may be affected by the possibility
of inducing multiple starbursts during a single merger event (Tissera 2000b).
Moreover, the formation of the stellar bulges of spiral galaxies can be directly
affected by the possibility of
combining two feasible formation mechanisms, such as secular evolution
and collisions, during a single merger event.
Observationally,  there are evidences that suggest that these different
processes, together with 
collapse at high $z$, could  work
with different relative importance, depending on the Hubble type, to
manufacture stellar bulges (Ellis, Abraham \& Dickinson 2001).
 
In this paper, we
 focus on the study of double starbursts
within the context of galaxy formation in a cosmological framework.
 Contrary to pre-prepared mergers, we use fully
self-consistent cosmological
hydrodynamical simulations, where the 
distribution of
merger parameters, such as orbital
characteristics, the orbital energy and angular momentum, the masses
of the virial haloes and baryonic clumps involved in the merger event, and the 
spin, internal structure, and relative orientation of the baryonic clumps
that are about to merge, among others, arise naturally at {\it each
epoch\/} as a consequence of the initial spectrum of the density fluctuation
field, its normalization, and the cosmological models and its parameters. 
The effects of mergers and
interactions on star formation can be studied in a consistent
scenario, although at the expense
of losing numerical resolution.
In Section 2 we present the  main aspects of the numerical models and
in Section 3, 4 and 5, the analysis of starbursts. 
 Section 6 summarizes the results.

\section{Numerical Models}

The simulations analyzed 
take into account the gravitational and hydrodynamical evolution of the matter
including an algorithm to transform the cold and dense gas into stars.
We use the Smooth Particle Hydrodynamical (SPH) 
code developed by Tissera, Lambas \& Abadi (1997)  coupled to
the AP3M (Thomas \& Couchman 1992).
SPH based codes have proved to be powerful tools to  study
galaxy formation 
(e.g., Navarro \& White 1994; Steinmetz \& Navarro 1999; DTTS98;
Tis00; Sal01).

We run four simulations, S.1, S.2, S.3 and S.4, 
 consistent with  Cold Dark Matter universes
with $\Omega =1$, $\Lambda=0$. The normalization of the power
spectrum was taken to be  $\sigma_8=0.40$ for S.1 and S.2 and
$\sigma_8=0.67$ for S.3 and S.4 (see Table~1).
The simulated boxes
have $5 h^{-1}$ Mpc length ($h=0.5$) with $N=64^3$ total particles.
 Baryonic particles represent 10 $\%$
of the total mass. Note that dark matter and baryonic particles have the same mass,
$M_{\rm part}= 2.6 \times 10^8 \ {\rm M{_\odot}}$.
The gravitational softening used is 3 kpc, and the smaller smoothing 
length allowed is 1.5 kpc.  
Simulations S.1 and S.2 share the same initial conditions (A)
while S.3 and S.4 are different realizations of  the power spectrum (B, C).
Run S.1 corresponds to experiments S.1 in DTTS98,  Sal01
 and Tis00.
Simulation S.2 has been run with a higher star formation efficiency
in order to assess the dependence of the results on this parameter.

The star formation (SF) algorithm transforms dense and cold gas
in a convergent flow 
into stars according to the   Schmidt law (Navarro \& White 1994; Tis00).
We assume a  certain  
star formation efficiency, $c$ (see Table~1),
and a characteristic time-scale
for the star formation process.
 The SF
parameters used in S.1, S.3 and S.4  allow the transformation of gas into stars only in the very dense regions.
In practice, this condition implies that gas particles have to 
fall in all the way down to the very central regions of GLOs
 before they
are converted into stars (see Fig.~2). This assures
the formation of a compact central stellar mass concentration,
that is, a bulge, and  that 
 the transformation of gas into stars on the discs is suppressed.
Conversely, the SF process in S.2 is more efficient,
 which implies that
gas particles are transformed into stars sooner than those in S.1. 
This difference translates in 
different physical properties for the final
objects, as Tissera \& Dom\'{\i}nguez-Tenreiro (1998) have shown.
These authors study the mass distribution of galactic objects in 
S.1 and S.2 finding that, because of the higher star formation
efficiency, those in S.2 tend to be dominated by
stellar spheroidal components with potential wells not
as deep as those of their counterparts in S.1.
The results of S.2 have been included  in this work to show that
the conclusions we reach here 
  are not mainly determined by
the star formation parameters, implying that the mechanism at work
is a general physical process.

We  want to point out that, because the SF process is based on 
the Schmidt law,  any sudden increase
of the gas density directly implies a corresponding
increase in the star formation activity. According to Tis00 (see also MH96),
it is possible to
have a large percentage of gas that satisfies the
SF criterion, but that is transformed into stars at a 
quiescent rate. And,  on the other hand,
 a smaller fraction of gas   can be  suddenly
compressed,  triggering a  starburst.

Supernova feedback effects have not been included in these simulations. Energy
feedback
processes are believed to play a key role in helping to set a self-regulated
star formation regimen (Silk 2001). However,
these processes are not well understood and 
 its modelling in hydrodynamical
simulations is still quite controversial (Katz 1992; Navarro \& White 1994;
Metzler \& Evrard 1994; Yepes et al.\ 1997;
Mosconi et al.\ 2001). And so,
as a first step, 
it is wise to analyze star formation  separately from supernova effects.  
Furthermore, 
as pointed out by
DTTS98 and Sal01, the
 restrictive SF  model used in this work
could have allowed us, in a way, to mimic
the effects of supernova energy feedback. 
                           In fact, the SF algorithm we have used
allows the formation
of stellar bulges without quickly depleting
the  gas reservoirs
of the objects. 
Sal01 have shown that  the formation 
of compact stellar bulges provides
 stability to the disc-like objects, so
that the disc component is  able to conserve an important fraction of 
its angular momentum during violent
events, producing  systems that resemble current normal  spirals. 
Particularly, these authors have analyzed the dynamical and
structural properties of GLOs in S.1 at $z=0$,
 finding that they have observational
counterparts. This is also valid for  those GLOs in S.3 and S.4 that
resemble disc-like objects. 
Hence, we are working with a set of objects that can be fairly
compared to spiral galaxies at $z=0$.\\

\section{Star Formation Histories and Mergers}

Galaxy-like objects  are identified at their
virial radius,
    $r_{200}$ (i.e., the radius where 
$\delta \rho /\rho \approx 200$; White \& Frenk 1991)
 at $z=0$. We reject those GLOs with a massive companion within two virial radii
at $z=0$ to avoid complications due to tidal fields originated by the presence
of companions and/or  the underlying overdensity, focusing
on the effects produced by the assembly of each individual object 
through hierarchical growth. 
We are only going to study
those GLOs with more than 250 baryonic particles inside the main object,
and, generally,  from $z=1$ to diminish as much as possible numerical resolution
effects, which could  be present at higher $z$.
 No other constraints have been imposed to select
the analyzed GLOs. 

A typical GLO is composed by a dark matter halo, a main
baryonic clump and a series of small baryonic satellites. Table~2 gives the
total number of dark matter ($N_{\rm dark}$) and baryonic ($N_{\rm bar}$)
particles within the virial radius for each GLO at $z=0$. 
Note that  dark matter haloes  are very well resolved
by several thousand particles. According to 
results from Steinmetz \& White (1997), if the potential well of a 
galactic halo is well-reproduced, the dissipative component
is forced to set onto the correct density profile, even if it is
resolved by few hundred particles. If the gas density profile
is adequately described, then the SF process within the GLO can
be reliably followed (see also DTTS98).

 The  merger trees of the GLOs  are reconstructed by
tracking back all particles that belong to a GLO
at $z=0$. The progenitor object is chosen as
the more massive clump within this merger tree.
A  merger event will be defined as the whole process from the instant
when the
 two main baryonic objects are first identified to share the same
 dark matter halo (this instant is the beginning of the ODP, whose
redshift will  hereafter be termed $z_{\rm o}$),
until the instant they  actually collide
at a redshift $z_{\rm c}$  (once the
ODP is over). Note that, in this paper, 
 a collision 
 means that the baryonic systems  {\it cannot be longer
separated\/} into two distinctive objects, but it does not imply any 
particular characteristics for the remnant.
 During the
ODP, the minor colliding baryonic clump
will be referred to as  the satellite.

 The  corresponding star formation rate (SFR) histories
of the progenitors 
can be  constructed  as the ratio between the 
stellar mass formed during a time-step of
integration and its duration ($\Delta t= 1.3 \times 10^{7}$ yrs for
S.1 and S.2 and $\Delta t= 1.2 \times 10^{7}$ yrs for S.3 and S.4).
These SFR histories  can be described as a result
of two contributions, a quiescent SF and a series of starbursts
 (see Tis00 for details). 
The quiescent SF gives rise to a constant SFR or threshold;
starbursts are defined to be those points of the SFR history
of a given object above its threshold (see Fig.~1).

Mergers of substructure are responsible of the triggering of some
of these SBs. 
>From all SFR peaks identified, in this paper we are only concerned with those that occur within  merger events. We have studied 29 merger events related to 
43 starbursts of which 14 are triggered during  orbital decay phases.
Once the starbursts are isolated, it is direct to define their maximum SFR, $\sigma_{\rm star}$, 
the total stellar mass formed, $M_{\rm burst}$, and their duration, $\tau_{\rm burst}$.
These quantities determine the characteristics of each starburst, together with the
gas richness of the system, $M_{\rm star}/M_{\rm bar}$, and the virial mass ratio of
the colliding  objects, $M_{\rm sat}/M_{\rm pro}$, at the time of the 
merger.
$M_{\rm star}$ and $M_{\rm bar}$ are the stellar and baryonic
 (i.e., gas plus stars) content of the
merging systems within $r_{200}$;
$M_{\rm sat}$ and $M_{\rm pro}$ are the  virial masses of the 
satellite and progenitor, respectively.
Table~2 gives their values for the mergers analyzed in this work.

\section{Double Starbursts}

During some merger events, two starbursts  
can be identified in the SFR histories of the GLOs. 
We classify as a double SB those two well-defined 
SFR peaks that could
be directly related to the same merger process. In other words, those 
two bursts that 
appear
 in the time interval from $z_{\rm o}$ (i.e., the instant when
the satellite enters the virial radius of
the progenitor) to the actual collision of the baryonic cores
at $z_{\rm c}$.
Only those bursts that strictly satisfy this condition were taken as probably
induced by the same merger. Whenever a doubt exists, we choose to eliminate
the merger from our statistics so that it does not contribute as either, a single burst or a double one. This situation, actually, occurs once.
As an example, we show in Fig.~1  the SFR
 for typical GLOs  that experience (a) a double
SB (GLO~3 in S.1, Fig.~1a) and (b) a single one 
(GLO~3 in S.3, second merger, Fig.~1b)   as a function
of lookback time ($\tau(z)=1-(1+z)^{-3/2}$).
We also plot in each of these Figures the distances from the progenitor 
to the satellite
in the time interval  from $z_{\rm o}$ to $z_{\rm c}$\footnote{
  It is worth mentioning that in some merger events, 
  during the orbiting of the satellite around the
  progenitor, more than one SBs can be detected.
  These cases are not studied in this paper.}.
In Fig.~1a, we see that the first burst occurs during
the orbital decay phase, while the second one is triggered  
during the collision  of the baryonic cores.
In Fig.~1b, only at the impact of the two baryonic clumps an enhancement
of the SF activity is detected.
Note that when the satellite enters the virial radius of the progenitor, in 
both cases (a, b), the progenitors are equally gas-rich as can be appreciated
from Table~2.

For each double SB, we measured the $\tau_{\rm burst}$, $ M_{\rm burst}$,
  $\sigma_{\rm star}$
and the time interval  $\Delta t_{\rm burst}$
 between the primary  ($\sigma_{\rm star}^{1}$) and
the secondary  ($\sigma_{\rm star}^{2}$) burst maxima,
($z_{\rm burst, 1} >
z_{\rm burst, 2}$;
$\Delta t_{\rm burst}$ was measured from the last SFR value of the primary burst to 
the first point of the secondary one, above the thresholds  
defining the bursts).
The temporal separations between
the bursts, $\Delta t_{\rm burst} \approx 5 \times 10^{8}-5 \times 10^{9} \ {\rm yrs}$, are consistent with the collapse times of satellites estimated
from dynamical friction considerations (Navarro, Frenk \& 
 White 1995; Hernquist
1989).
Table~2 summarizes the results (secondary components of
double bursts are indicated by the letter D).
Last column on Table~2 gives the redshifts ($z$) at which each starburst starts.

 It is clear that if 
 there is   enough  available gas in the system,
a starburst will be triggered when the progenitor and the
satellite collide at $z_{\rm c}$.
Regarding double SB events, 
our concern now is to prove that the  SBs detected
in the progenitors  during some 
ODPs are produced by gas inflows driven as the satellite 
gets closer,   as predicted by  pre-prepared simulations (e.g., Hernquist 1989;
MH96).
Also, we want now to assess the differences, prior to the merger event,
of the internal structure of GLOs involved in double and single
SBs, as well as the evolution of these differences, in relation 
with their SFR histories, during and immediately  after the merger event.   
To this end, we have studied and compared the properties
of our GLO sample at the merger events reported in Table 2.
For each of these mergers, the variations of
the baryonic mass  and angular momentum 
distribution at the central regions of
 their respective progenitors have been followed and quantified,
in the time
interval between the beginning of their ODPs at $z_{\rm o}$ and
a reference time at $z_{\rm f}$\footnote{
  For mergers producing double starbursts, $z_{\rm f}$ has been taken to
  be the moment when the ODP ends, $z_{\rm o,f}$,
  because $[z_{\rm o}, z_{\rm o,f}]$ is the
  time interval within which the primary component of double starbursts is
  detected. For mergers that do not produce double starbursts, 
  we have taken $z_{\rm f} = z_{\rm c}$, that is,
  the time when the satellite and the
  progenitor actually collide.},
using different estimators.  

As a first step, 
 the  progenitor particle number (actually, the baryon mass) inside
a given radius, $N( < r)$, has been measured at different 
redshifts in the $[z_{\rm o}, z_{\rm f}]$ time interval.  
The general result is that gas inflows are detected during this
time interval at the central regions of  progenitors
that experience double starbursts. In contrast with this
result, no gas inflows are detected at any stage at the
central regions of progenitors involved in merger events
with single starbursts.
To illustrate these results,
we have chosen as typical examples the mergers occurring in GLO~2 from S.4 
(a double SB event at $z\approx 1.20 $)  and in  GLO~5 from S.1 
(the latest one at $z\approx 0.19$, a single SB event; see Table~2),
but the behaviour
is similar for the other merger events. 
In Fig.~2a we plot $N(<r)$ for  the  primary
 component of the double SB in GLO~2 from
$z_{\rm o}=1.81$ to  $z_{\rm f}=1.28$. 
Note how  gas particles are displaced from  the outer to the  inner
radius as the time goes on.
 Part of the particles that get into the 10~kpc $<r<$ 30~kpc
shell are transported
all the way down to the very centre ($r<$ 4~kpc). This increase in the gas density immediately translates into an 
increase of the star formation activity, as previously discussed.
The inward  displacement of these particles is due to a loss of their spin angular momentum 
content as the satellite gets closer (see DTTS98).
GLO~5 has assembled most of its mass at higher $z$, so that when the last
merger occurs,
 its gaseous disc and stellar bulge are clearly at place.
Fig.~2b shows $N(<r)$ between $z_{\rm o}$ and $z_{\rm f}$ for the last
merger. As can be seen, the changes in the baryonic mass
distribution are much less important in this case than for GLO~2.
There is almost no increase in the baryonic mass within 15 kpc or the inner 
4 kpc, implying
that no significant  gas inflows have been triggered in the central region during the orbital 
decay phase of this satellite.

In the other six plots of Fig.~2, we show
the angular momentum component per unit mass, $j_{z,i}$,
along the direction of the 
total angular momentum of the main baryonic systems,
for each gas and star particle in GLO~2 (left panels) and
 GLO~5 (right panels),
versus particle radial distance, $r_i$,
at three stages of evolution within the
interval of interest $[z_{\rm o},z_{\rm f}]$. These plots
give a clear picture of the internal structure of both GLOs during 
 such interval.
These Figures show that when GLOs have  well-formed stellar bulges,
the disc components are not significantly affected during the ODP. Conversely,
those GLOs which experience early gas inflows do so at the expense of
the gas component on the discs. These gas inflows feed the stellar
bulges. Fig.~2 also suggests that in the latter cases, GLOs
are in early stages of evolution.

To go further into  the quantification of
  the differences between the properties of the
GLOs at a double or single SB event,
 we  use  results
from 
  bulge-disc  decomposition of their  projected  mass surface  
density 
(Scannapieco et al.,  in
preparation).
 Following  Sal01, the 
 projected mass surface density  has been decomposed by
using a S\'ersic law for the bulge component and an exponential law for
the disc component.
 Sal01 have  shown  
that, at $z=0$,  this procedure
yields scalelengths and dynamical parameters which are in very 
good agreement with recent observations of spiral galaxies (e.g.,
Courteau 1996). 
Here, the aim of this decomposition is
to provide  structural parameters which  allow a
convenient  quantification 
of 
the mass distributions of progenitors
at $z_{\rm o}$ and of their changes between $z_{\rm o}$ and $z_{\rm f}$.
The analysis has been performed for GLOs in 
S.1 and S.4, for which a suitable number of outputs to follow their
evolution in time are available. The general  result is that  
the mean values of the 
 masses of the bulges at $z_{\rm o}$,
 as well as their variations  between $z_{\rm o}$ and $z_{\rm o,f}$, 
show clear  differences 
for GLOs involved in double or single starburst events. 
At $z_{\rm o}$, the former show 
smaller bulges than the later on average (i.e., $\approx 50\%$ smaller),
 but, after the  starbursts 
that take place within the  $[z_{\rm o},z_{\rm o,f}]$ interval, 
their bulges have grown 
by  $47  \pm 19\%  $ on average. Meanwhile,
those GLOs with no important SF during an equivalent time-interval, 
show  a change in the bulge mass of  $5 \pm 2 \% $\footnote{
  The errors correspond to standard deviations calculated by using
  the re-sampling bootstrap technique with 500 random samples.}.
These results confirm that  systems experiencing early gas inflows 
have smaller bulges which are fed by gas inflows during
the ODPs.

\section{Starburst Properties}
 
Taking into account previous works on disc formation, 
there are evidences that support the idea that the
stability of the disc structures plays
a relevant role in the triggering of gas inflows when the
systems are subject to tidal fields. This stability
could be assured by a compact stellar bulge  when
the dark matter (DM) halo is not dynamically dominating at the centre
 (MH96; Barnes \& Hernquist 1996).
 Then, a first step towards the understanding of the
physical origin of the induced SBs in our simulations,
  is to analyze any  possible relation between the formation
of the stellar bulge in the progenitor and the burst properties.

To this end, our first task is to define  the epoch of  bulge formation.
This  is a controversial point since
bulges can continuously grow by mergers,  
 bar-driven infall or other mechanisms
(Combes 2000).
The simplest possibility to  define the presence of a bulge
is to adopt   a threshold criterion relative to the bulge mass.
However, we want to employ a criterion which allows the
comparison among bulges of different virial masses.
Consequently, due to the correlation among bulge masses and
virial masses, a threshold criterion relative to 
{\it normalized\/} bulge masses should be used instead.
Moreover, the normalizing factor has also to be
sensitive to the history of evolution of  GLOs 
(i.e., their respective merger trees and interaction events),
because this history also affects the SF history of each GLO.  
We
found that the total stellar mass in the GLO at $z=0$ is sensitive
to both the virial mass of the object and its history of
 evolution.
Then, we define the ratio between the stellar mass formed
in the progenitor
 at a given $z$, 
$M_{\rm star}^{z}$, and the total final stellar mass, $M_{\rm star}^{z=0}$,
 of the GLO,  as an estimator of the stellar core
(i.e., bulge) formation
(see Fig.~2 for a general picture of baryon distributions). 
The monitoring of this ratio along the history of evolution of a GLO
allows us to adopt a clean criterion for the formation of bulges.

Now, if our hypothesis about the ability
of well-formed bulges to provide stability to a bulge-disc
system is correct, then there should be a correlation between
the $M_{\rm star}^{z}/M_{\rm star}^{z=0}$ ratios,
measured at a redshift $z = z_{\rm o}$, 
when the satellite enters the virial radius of the
parent object,  
 and the properties
of the starbursts.
For each merger in Table~2, we have
 calculated $M_{\rm star}^{z}/M_{\rm star}^{z=0}$ and
the relative strengths of the two bursts, $\sigma_{\rm star}^{1}/
\sigma_{\rm star}^{2}$. The results are plotted in Fig.~3.
As can be seen from this Figure, there is a clear segregation which indicates
that 
 when
a single burst ($\sigma_{\rm star}^{1}/
\sigma_{\rm star}^{2}=0$) is associated with a merger, 
GLOs tend to have larger $M_{\rm star}^{z}/M_{\rm star}^{z=0}$ values.
Conversely,
when 
a double burst is found, all of them have  $M_{\rm star}^{z}/M_{\rm star}^{z=0} \leq 0.40$.
Our results show that as the stellar bulge forms, the
relative strength relation reverses ($\sigma_{\rm star}^{1}/\sigma_{\rm star}^{2}< 1$)
 until the first burst disappears when the stellar bulge is
fully present  ($\sigma_{\rm star}^{1}/\sigma_{\rm star}^{2}=0$).
The ratio $M_{\rm star}^{z}/M_{\rm star}^{z=0}\approx 0.40$ seems to
indicate when the stellar bulges have grown large enough to
prevent even mild inflows.

Hence, it seems plausible that, depending on the structural properties of the
GLOs, the merger with a satellite may induce two starbursts whose
properties would also depend  on the internal structure of the galactic
objects. 
Our findings also agree with Hernquist (1989) results,  which 
suggest  that tidal triggering of starbursts
may be a natural way of explaining the discrepancy found between the inferred 
starburst time-scales ($\approx 10^{8}\rm yr$)
 and the dynamical time-scales of the objects ($\approx 10^{9}\rm yr$) .

\subsection{Dependence on the potential well}

In order to intend to relate the characteristics of the
potential well in the very central regions of GLOs 
with the presence of a well-formed bulge as measured
by  $M_{\rm star}^{z}/M_{\rm star}^{z=0}$,
 we
estimate the circular velocity  at $r=$ 2~kpc, $\upsilon_k$,
for the dark matter ($k=$ DM) and  gaseous  ($k=$ gas) components and
the total mass ($k=$ tot) at $z_{\rm o}$.
 This is a simple but straightforward way of measuring the
mass concentration.
For this purpose, we calculate the circular velocity curves defined as
$V(r)^2=GM(r)/r$,
for the three mass components at primary components and single bursts.
We use velocities instead of density profiles
since the former are defined by the  integrated mass which is less
affected by  numerical noise. Note also that for this calculation we
are not including the gravitational softening since we are interested
in measuring the actual mass distribution. However, the correct
circular velocity curves should be estimated by taking into account
such softening  as shown by  Tissera \& Dom\'{\i}nguez-Tenreiro (1998)
and  Sal01. 

Figs.~4~(a,b,c)  show $\upsilon_k$ for the total mass, DM and
gaseous components, respectively,
 against the
ratio between the heights of the starbursts 
$\sigma^1_{\rm star}/\sigma^2_{\rm star}$ (recall that
for single bursts, $\sigma^1_{\rm star}/\sigma^2_{\rm star}=0$) 
for bursts in S.1 and S.4.
The segregation is quite clear, indicating that
only when the systems have  less concentrated potential wells measured
 by lower $\upsilon_k$ values,
gas inflows are induced,
triggering starbursts during the orbital decay period. Note that
all mass components show a similar trend.
As it can be appreciated from Fig.~3,  systems with 
$\sigma^1_{\rm star}/\sigma^2_{\rm star} \neq 0$
  have smaller stellar bulges.
 Because of the new gas infall, the stellar bulges  grow
 as the gas is transformed
into stars at the centre. When GLOs have well-formed bulges,
they  have already  pulled in enough mass to
assure the stability of the system which prevent further important inflows.

\subsection{Starburst efficiencies}

>From Fig.~3  we showed that double bursts occur only  when
 $M_{\rm star}^{z}/M_{\rm star}^{z=0} \leq 0.40$.
It was found by Tis00 that the efficiency in the transformation of gas into stars during  SBs    
 correlates with $M_{\rm sat}/M_{\rm pro}$  in the sense that
 massive mergers ($M_{\rm sat}/M_{\rm pro}\geq 0.40$) seem to be more efficient at converting gas into stars.
Here we  explore the possibility that  the
SB efficiencies (defined as the ratio  between the stellar mass formed during
a burst and the total gas mass available in the system at the time of the merger, $M_{\rm burst}/M_{\rm gas}$)
 of the bursts triggered during the ODP  may be  different
from those of SBs produced at the actual collision of the baryonic clumps.
In double SBs we do not consider the secondary
components, since they form from the  left-over gas in the 
progenitor after the primary one and the gas component associated to the satellite.
They are also difficult to study since we would need to temporary resolve
the merger with a higher number of outputs 
and higher numerical resolution.

Fig.~5 shows $M_{\rm burst}/M_{\rm gas}$ vs
 $M_{\rm star}^{z}/M_{\rm star}^{z=0}$. Among GLOs in the same simulation,
the larger values are mainly recorded for the smaller
$M_{\rm star}^{z}/M_{\rm star}^{z=0}$.
This can be clearly appreciated from Table~3 which shows the  mean values 
of $M_{\rm burst}/M_{\rm gas}$ for  $M_{\rm star}^{z}/M_{\rm star}^{z=0}$  smaller
or greater than 0.40.
Note that mean starburst efficiencies are higher for GLOs  in S.2 simulation.

Hence, we find a clear trend for SBs induced during the ODP to be 
more efficient at converting gas into stars
 than those produced by the actual collision of the
baryonic cores. This result may have strong implications
for galaxy formation.

\subsection{Dependence on redshift}

In Figs.~6a~and~6b we plot   $M_{\rm sat}/M_{\rm pro}$ and
$M_{\rm star}^{z}/M_{\rm star}^{z=0}$, respectively, 
as a function of the redshift when the starbursts start ($z_{\rm burst}$ on Table~2).
We include single bursts  and primary components of double SBs (features with
arrows).
 As can be seen form Fig.~6a, there
is a clear trend to have major mergers (i.e., similar mass objects)
at high $z$ as expected in a hierarchical clustering
model (e.g., Lacey \& Cole 1993).
We also see that  double SBs tend to occur at higher $z$, and
consequently, to be linked to more important merger events,  with 
$M_{\rm sat}/M_{\rm pro} \geq 0.20$ for all double events.
In Fig.~6b, $M_{\rm star}^{z}/M_{\rm star}^{z=0}$ as a function of $z _{\rm burst}$
 shows how, 
as the progenitors build up their
stellar bulges  as they evolve, the probability of 
experiencing   bursts induced via tidal forces diminishes.
Note that, in these models, all double events occur at $z>0.4$.

>From the combination of the information  coming out from Fig.~6,
we note that $61 \%$ of  bursts induced during the ODP 
are triggered by minor mergers ($M_{\rm sat}/M_{\rm pro}
\leq 0.40$). Even more, if we consider only SBs in S.3 and S.4, which
constitute an homogeneous sample (i.e.,  different
random phases of the power spectrum run with the same model parameters),
the trend is very similar: $66\%$ of the SBs detected during the 
ODP are related to minor mergers.
 This tendency suggests that the relative mass
of the colliding objects is not the determining mechanism
in the triggering of the early gas inflows.
On the other hand,
none of the  systems
with well-formed stellar bulges, $M_{\rm star}^{z}/M_{\rm star}^{z=0} > 0.40$,
 have starbursts induced during the ODPs (Fig.~3).

>From these results, we conclude that 
hierarchical clustering is determining the global trend
 for massive mergers to occur at higher $z$,
when the systems are more likely to be unstable, and, consequently, 
more susceptible to experiencing
gas inflows and double SB events.
We predict that this trend to have SBs induced via tidal forces at higher $z$,
combined with their higher efficiency in transforming gas into stars, lead
to a global star formation efficiency that would increase with $z$. 
Observational analysis of the star formation efficiency with $z$ should
test this prediction.

\subsection{Bulge formation}

Three different formation scenarios have been proposed for
bulges of spiral galaxies: dissipative collapse (Eggen, Lynden-Bell \&
Sandage 1962), mergers of previously formed systems (e.g., 
Kauffmann, White \& Guiderdoni 1993)
and secular evolution (see Combes 2000 for a review).
 It has been also suggested that the three of
them may have played a role,
 but with different relative efficiencies depending
on the Hubble type (e.g., Combes 2000; Ellis et al.\ 2001).
Our results  support such hypotheses.
Moreover, they  show that a given 
merger event could contribute to the formation of stellar bulges
through two different processes: a) 
 gas inflows  triggered during  interactions at the ODPs, and b)
collision of baryonic cores, 
once the ODPs are over.
The first contribute to the secular evolution of bulges and
have been  found to be more important at higher $z$, as previously
discussed (see Figs.~6;  
note that these inflows
are episodic). Together with  this secular  component, there is a collisional
contribution produced during the actual mergers of the main
 baryonic clumps, so that after a merger event, the stellar 
population of the GLO involved in the merger
consists of 
the contribution of both components, plus the old stars.
The relative importance of each of these stellar 
 components would  depend on  the
previous evolutionary paths of the objects and
their physical internal characteristics  at the time of each merger,
particularly on those of their potential wells (see subsection 5.1).

According to this scenario, the stellar bulges will have stellar
populations with different ages. For the particular models
analyzed in this paper, $50 \%$ of the stars in the bulge are formed
at $z \ge 0.5$. If we assume that, morphologically, the bulge component
is at place when $50\%$ of its total stellar mass is formed, then
$z\approx 0.5$ will be its average age in these CDM scenarios.
 However, if the older stellar 
populations are taken to be  age indicators, then the time of
formation is displaced to much higher $z$.

The time of formation of the bulge component could be affected by the
cosmological model as well as by the treatment of supernova effects.
In particular, the lower normalization parameters ($\sigma_8$) adopted for
these simulations  contribute to produce a late formation of 
bulges. In a higher  normalization scenario, mergers would occur at
higher $z$,  with an earlier formation of central mass concentrations.
Note also that higher numerical resolution can push the age towards
higher redshift. Hence, our age indicators should be taken as
lower limits. Nevertheless, the physical processes and, principally, the
fact that stellar bulges are found to be formed naturally as the 
product of collapse, collisions and secular evolution, remain valid.
Although previous works proposed such mechanisms  to work together in the
manufacturing of stellar bulges,  
the results of the simulations reported here
constitute the first proof  
 that these mechanisms are  actually
 at work  in a cosmological framework and,
also, they give us a preliminary information on  how they do take place.
The study  of bulge formation by means of
  numerical simulations is very convenient,  since
it can provide  hints on the physical processes 
involved and their interplay, while 
in semi-analytical models they have  to be modelled following recipes 
known in advance (e.g., van den Bosch 2001).

Observationally, it has been recently shown by Ellis et al.\ (2001) that
at $z \le 0.6$ the bulges of spiral galaxies are bluer than ellipticals,
 suggesting that the stellar population has been rejuvenated, specially at
intermediate $z$. This has been suggested to be in contradiction
with predictions of 
some  hierarchical clustering scenarios made by
semi-analytical models,  where spiral bulges and ellipticals are found to
have very similar characteristics,
as, in these models,
 their stellar populations have formed mostly at collapse time.
However, our results show that hierarchical clustering scenarios
can be reconciled with these observations,
 since  the bulge stellar populations formed through mergers
events,  might be the rejuvenating  agent after core collapse.

\section{Discussion and Conclusions}

The main advantage of the method reported in this paper
to study mergers, over pre-prepared simulations, is that
we analyze 
mergers that occur as the result of the self-consistent formation
of the structure within a cosmological model. Hence, the physical
properties  of the colliding objects and the  parameters of
the encounters are
those that correspond to galactic objects evolving in the adopted
framework, without any external manipulation.

In particular, the processes of gas infall at scales of hundreds
of kpcs and  gas inflows in discs, at smaller scales,
 appear in a natural way
in our simulations. Both of these processes  are
 a consequence of  evolution  at larger scales,  from
initial conditions that sample a given spectrum of
primordial fluctuations (CDM in this case) at very high $z$.
The importance of these processes in the
SFR history of GLOs lies in the role they play,
not only as suppliers of gas to be turned into stars
in GLOs, but also in the fact that they provide
the baryon mass  to form a central compact concentration,
that, later on, makes it  difficult any  further gas inflow.
It is this possibility of analyzing  different phenomena
involved in galaxy formation, and the dynamical and
hydrodynamical connections among them, that renders attractive
the self-consistent approach to study mergers reported in
this article.

Our results suggest that the internal structure
of the GLOs  affects its SFR history, as reported by MH96 among others.
Because  a disc structure is a natural attractor for the dissipative
component, its stability is crucial for its later evolution.
The possible formation of  radial inflows which can re-distribute
the gas within the central regions and accumulate it in very short time-scales,
directly affects the SFR in the systems.

In agreement with previous works, we  find that some mergers
 can induce star formation in two stages, depending
on the internal  mass distributions of the parent galaxy.
The second burst is simultaneous with the collision of the baryonic
cores, while the first one occurs during the orbital decay phase of
the satellite.
We also found that the starburst efficiency of the primary bursts
tends to be larger than that of the secondary ones, implying that induced
starbursts during the orbital decay phase can be more efficient at transforming
gas into stars than those produced in the actual collision of the baryonic cores.
Therefore,
although the rate of gas cooling and accretion onto the
main baryonic clump,
 at scales of say $    \simeq  $ 100 kpc,
 can be mainly determined by the dark matter halo (White \& Frenk 1991),
how this gas distributes itself and responses to
external and/or internal factors within the central  GLO regions
is a small scale process  and it is affected
by its particular evolutionary history
(DTTS98; Sal01).

A specific result of our method of analysis is  the finding that
a self-regulated process among gas, stars and dark matter components is
settled. This self-regulation is a basic clue to understand bulge formation
that can only be studied in a self-consistent scenario.
Our results show that  single SBs tend to occur in systems with  
larger $M_{\rm star}^{z}/M_{\rm star}^{z=0}$,
which is pointing out the fact that these systems have suffered gas inflows  in
previous phases of evolution which have fed their bulges.
As a consequence, the dark matter has also got
more concentrated as the response to the baryon concentration (see Tissera
\& Dom\'{\i}nguez-Tenreiro 1998).
Conversely, systems with
 shallower potential wells tend to be  in the process of forming the bulge.
During this process,
the whole potential well gets deeper.
The three components, gas, stars and DM, work together.
 It is the response to
their coupled evolution that can be correlated with the effects a merger event
may cause on  the dynamics of
the systems and, as a consequence, on their star formation histories.
Other characteristics of mergers,  such as  orbital orientation,
 may also have
influence on the response of the systems as reported by 
Barnes \& Hernquist (1996).
Their detailed study within a hierarchical scenario
will be carried out in a separate paper.

We  found that in a hierarchical clustering scenario, there  is a natural
evolutionary sequence:   mergers
 occurring at higher $z$ when the systems
are not that concentrated and stellar bulges not well-formed, lead
 to gas inflows that
trigger SBs and feed the first bulges.
 As a consequence, the
total  potential wells concentrate.
When the objects reach lower $z$, they tend to be more stable because of
their  important stellar bulges and
concentrated  potential wells that prevent further  gas inflows.
At this stage, mergers also tend to
be minor.
This evolutionary sequence suggests 
that starbursts induced during the ODP are likely to be more  common at
higher redshifts; and because of their higher efficiency in
transforming gas into stars, we predict that their impact on galaxy formation
could be  very important, particularly, on the formation of spiral bulges.

Finally, our results suggest that hierarchical clustering scenarios
can be reconciled with recent observational results of high redshift
spiral bulges and ellipticals in the field, since,  in 
such  scenarios, stellar bulges would form by
the  contribution of collapse, collisions and secular evolution,
each one with a relative importance determined by the 
evolutionary path of each system.

We are grateful to
 the University of Oxford (United Kingdom),
 Imperial College of Science, Technology and Medicine (United Kingdom), Observatorio Astron\'omico de 
C\'ordoba (Argentina) and Centro de Computaci\'on Cient\'{\i}fica de la 
Universidad Aut\'onoma de Madrid (Spain)
for providing the computational support for this work.
This work was partially supported by MCyT (Spain), through grant
 \mbox{AYA-0973}, Consejo Nacional de Ciencia y Tecnolog\'{\i}a 
(Argentina) and Agencia 
de Promoci\'on Cient\'{\i}fica y T\'ecnica (Argentina). 


\clearpage

\begin{table}
\caption{Numerical Simulations}
\begin{tabular}{cccc}
S& CI & $\sigma_8$  & $c$ \\
S.1& A& 0.40&0.01\\
S.2&A&0.40&0.10\\
S.3&B&0.67&0.01\\
S.4&C&0.67&0.01\\
\end{tabular}
\end{table}

\begin{table}
\caption{Main Characteristics of Starbursts}
\label{main}
\begin{tabular}{cccccccccccc}
S & 
GLO   &
$N_{\rm dark}$ & $N_{\rm bar}$& 
$\sigma_{\rm star}$   &
$M_{\rm burst}$ &
$\tau_{\rm burst}$ &
 $M_{\rm sat}
/M_{\rm pro}$&
$M_{\rm star}/M_{\rm bar}$ &
$M_{\rm star}^{z}/M_{\rm star}^{z=0
}$&
$\Delta t_{\rm burst}$& $z_{\rm burst}$\\
\hline
S.1&1&9181&1335&18.75&0.55&4.36&  0.16& 0.12&0.20&&0.50\\
& &&&33.01&4.06&19.53&  0.23& 0.10&0.30&27.09&0.41\\
& &&&15.47&1.12&8.61& D&& &&0.15\\
&2&7310&1059&24.76&0.96&3.53&  1.08& 0.14&0.10&5.42&0.63\\
& &&&16.52&0.10&1.99& D & &&&0.54\\
& &&&46.43&4.63&17.60&  0.39& 0.26&0.35&&0.47\\
& &&&12.38&0.26&5.76& D & &&&0.29\\
& &&&17.53&1.25&10.11&  0.11& 0.28&0.90&&0.08\\
&$3^*$&6317&917&13.41&0.91&11.50&  0.40& 0.22&0.40&16.67&0.56\\
& &&&28.89&2.18&11.21& D& &&&0.34\\
&$4^*$&7115&908&29.92&1.44&13.50&  0.16& 0.24&0.60&&0.34\\
& &&&12.38&0.65&10.39&  0.16& 0.27 &0.85&&0.14\\
&$5^*$&5215&739&8.25&0.49&2.48&   0.43& 0.27&0.42&25.01&0.41\\
& &&&19.60&0.89&12.4& D&&&&0.34\\
& &&&22.70&0.91&5.14&  0.22 &0.21&0.75&&0.19\\
&$6^*$&6184&798&16.50&1.71&11.50&0.40&0.15&0.25&2.58&0.78\\
&&&&17.53&1.09&8.11& D&&&&0.52\\
S.2&1&9199&1307&42.30&0.68&3.94& 0.18 &0.58&0.20&&0.56\\
& &&&43.33&2.76&13.40& 0.22& 0.61&0.63&&0.43\\
&2&7383&1071&52.62&1.56&8.86& 1.06& 0.51&0.15&13.01&0.81\\
& &&&30.96&1.72&6.01& D&&&&0.56\\
& &&&71.19&4.18&19.08& 0.38& 0.54&0.40&&0.39\\
& &&&66.93&0.36&8.13& 0.11& 0.59&0.95&&0.09\\
&3&6501&918& 30.95&1.56&11.21& 0.43& 0.62&0.90&&0.34\\
&4&7144&921&39.21&1.87&7.47& 0.10& 0.66&0.55&&0.37\\
& &&&41.12&1.53&13.99& 0.21& 0.46&0.88&&0.13\\
&5&5294&744&40.24&2.11&15.03& 0.51& 0.78&0.45&15.80&0.49\\
& &&&33.02&1.43&19.30& D& &&&0.30\\
S.3&1&3688&550&23.73&0.57&0.81& 0.31& 0.10&0.30&37.00&1.47\\
& &&&5.16&0.05 &1.23& D&&&&0.58\\
&2&5665&1056&8.25&0.60&14.25& 0.58& 0.15&0.34&31.11&0.65\\
 &&&&5.16&0.44&13.38& D& &&&0.25\\
&3&1168&272&22.69&0.43&2.89& 0.35& 0.11&0.25&&0.98\\
& &&&6.19&0.18&3.12& 0.10& 0.21&0.80&&0.09\\
&4&3302&567&18.57&0.47&2.70& 0.24& 0.26&0.01&0.32&5.46\\
& && &10.32&0.36&2.76& D& &&&3.96\\
S.4&1&6541&646&17.17&0.83&8.76  & 0.57 &0.20  &0.21&2.39&0.87\\
& &&&23.24&1.46  &9.97  & D &  &&&0.65\\
&2&5761&621&16.27&1.01  & 9.88 & 0.28 & 0.18 &0.18&0&1.40\\
& &&&17.88&1.20  &12.33  & D &  &&&1.07\\
&3&7976&1143&22.59&0.83  &6.17  & 0.23 &0.30  &0.11&3.70&2.05\\
& &&&9.94&0.07  &10.01  & D &  &&&1.51\\
&&&&10.41&0.31  & 6.08 & 0.14 &0.54  &0.71&&0.34\\

\end{tabular}

\medskip

Units:
$[\sigma_{\rm star}] = {\rm M_{\odot}/yr}$;
$[M_{\rm burst}]=10^{10}\,{\rm M_{\odot}}$;
$[\tau_{\rm burst}]=10^{8}\,{\rm yr}$;
$[\Delta t_{\rm burst}]=10^{8}\,{\rm yr}$

\end{table}

\begin{table}
\caption{Mean Starburst Efficiencies}
\begin{tabular}{ccccc}
S& S.1 & S.2  & S.3 & S.4\\
$M_{\rm star}^z/M_{\rm star}^{z=0}\leq 0.40$&0.39&0.56&0.16&0.11\\
$M_{\rm star}^z/M_{\rm star}^{z=0}>0.40$&0.08&0.39&0.03&0.03\\
\end{tabular}
\end{table}

\clearpage

\section*{Figure captions}

\clearpage

\begin{figure}
{\psfig{file=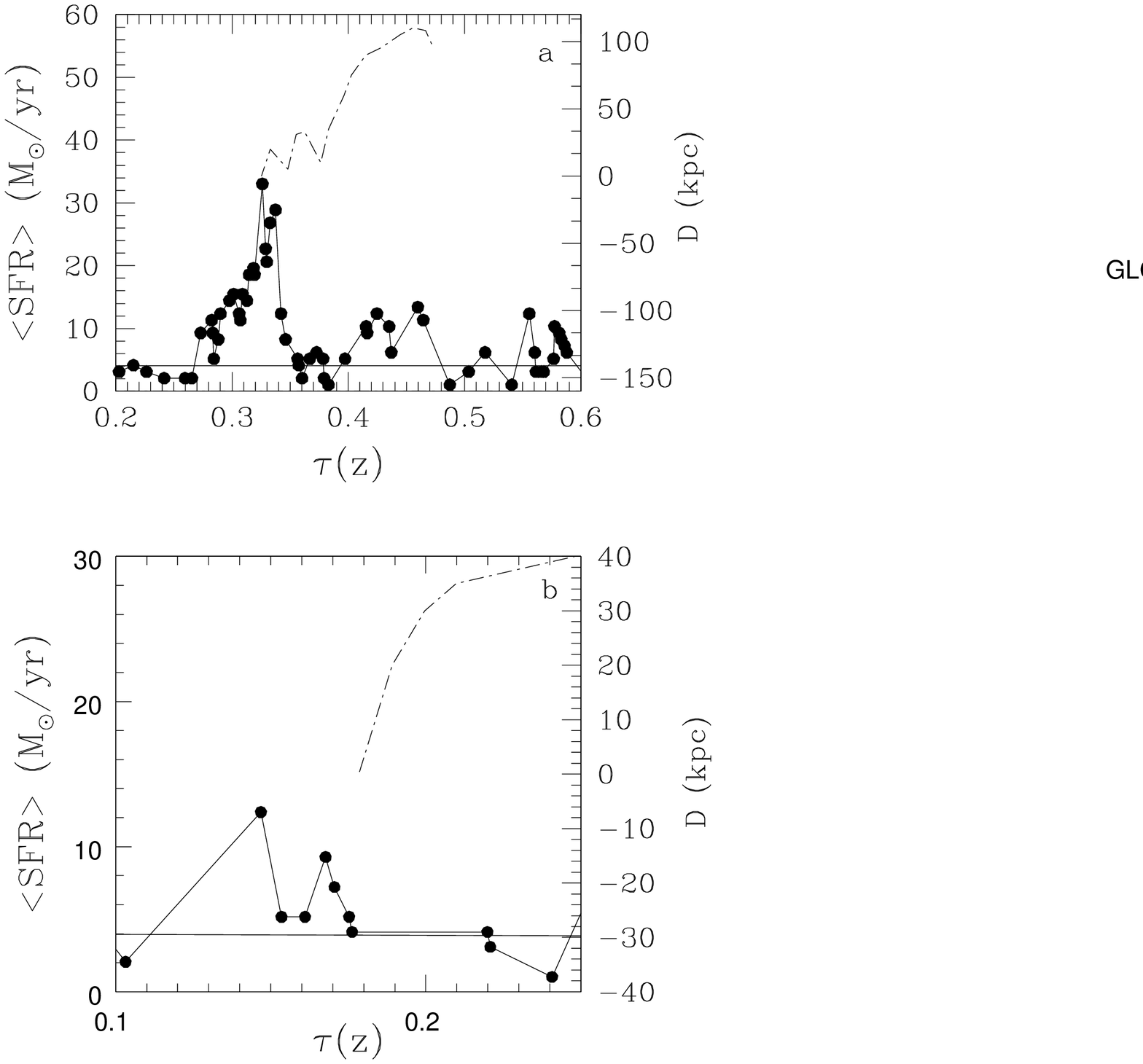}}
\caption{
 Star formation rate for a typical galaxy-like object
 experiencing a double starburst (a) and a single one (b) 
in simulation S.1 as a function of lookback time. The spatial separation between
the mass centres of the progenitor and satellite clumps are
plotted ({\it dashed lines\/}).  The {\it solid lines\/} represent the
quiescent star formation component. }
\end{figure}

\clearpage

\begin{figure}
{\psfig{file=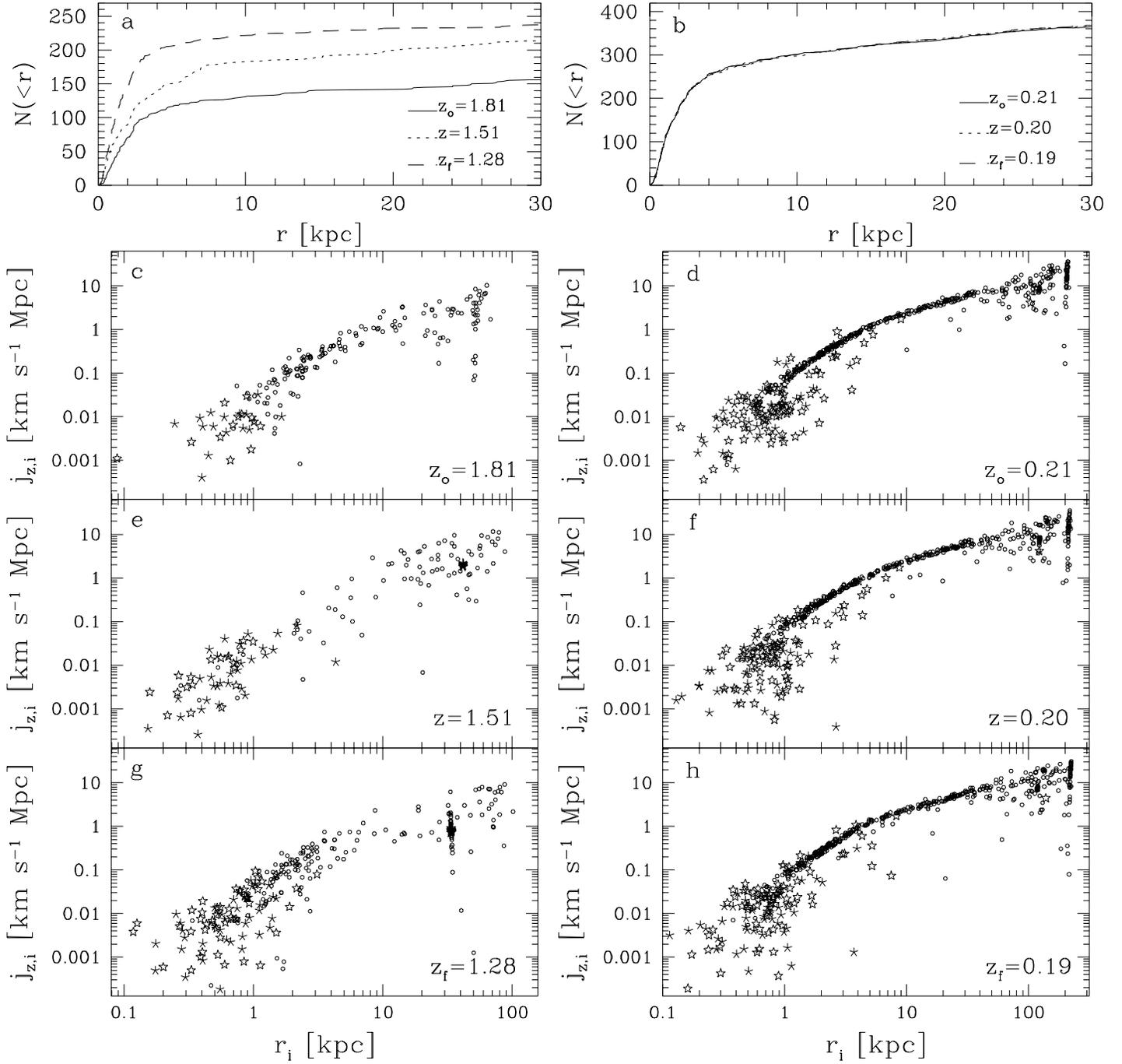}}
\caption{
 Number of baryonic particles 
within radial distance $r$ for  GLO~2 (a) in S.4 and  GLO~5 (b)  
in S.1
 between $z_{\rm o}$ and $z_{\rm f} $.
Specific angular momentum component  
$j_{z,i}$
along the direction
of the total angular momentum of the progenitor
for each baryonic particle within $\approx$ 100 kpc 
 of GLO~2 (c,e,g) and  GLO~5 (d,f,h) versus their positions
$r_i$
from   $z_{\rm o} $  to $z_{\rm f} $.
{\it Circles\/}: gas particles, {\it  stars\/}: 
star particles, {\it asterisks\/}: counterrotating stars. }
\end{figure}

\clearpage

\begin{figure}
{\psfig{file=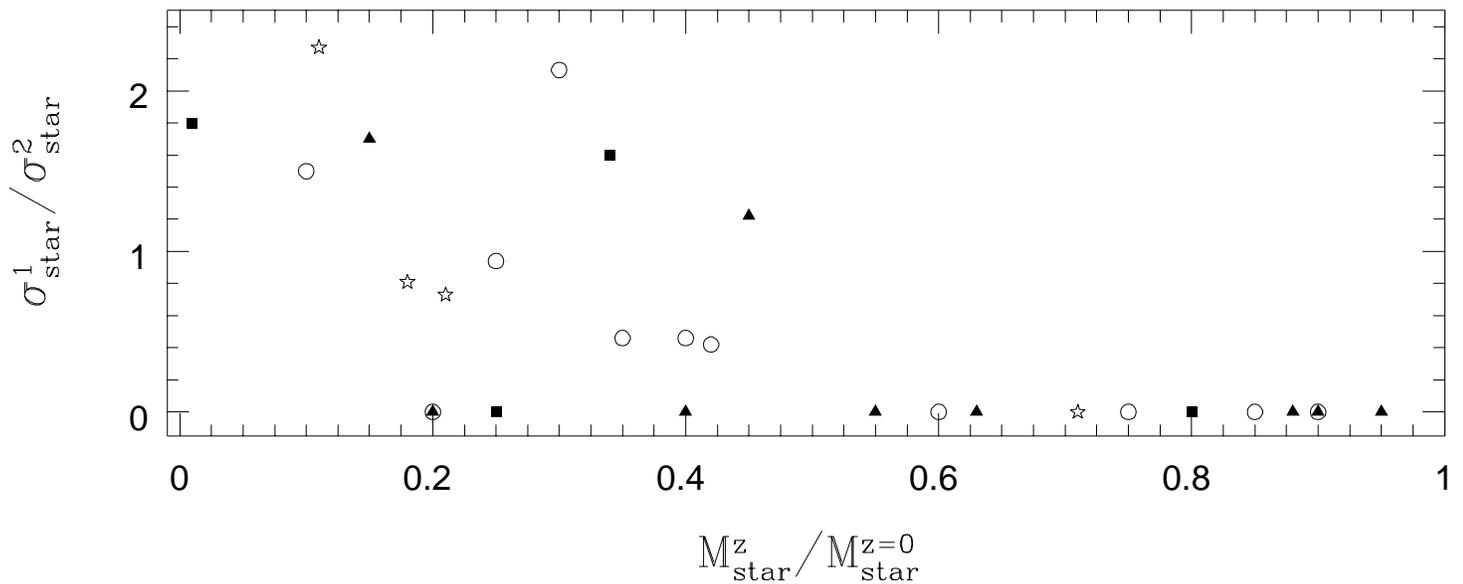}}
\caption{
 Ratio between the strengths of the
double starbursts ($\sigma_{\rm star}^{1}/\sigma_{\rm star}^{2}$)
and the fraction of stars already formed  in the
progenitor  ($M_{\rm star}^{z}/M_{\rm star}^{0}$)
at the $z$ of the merger.
Single bursts have been assigned
$\sigma_{\rm star}^{1}/\sigma_{\rm star}^{2} =0$.
GLO~1 in S.3 has not been included for the sake of clarity
($\sigma_{\rm star}^{1}/\sigma_{\rm star}^{2} =4.6$).
 GLOs in S.1: {\it open circles\/}, S.2: {\it filled triangles\/} S.3: {\it filled squares\/} and S.4: {\it open stars\/}.
}
\end{figure}

\clearpage

\begin{figure}
{\psfig{file=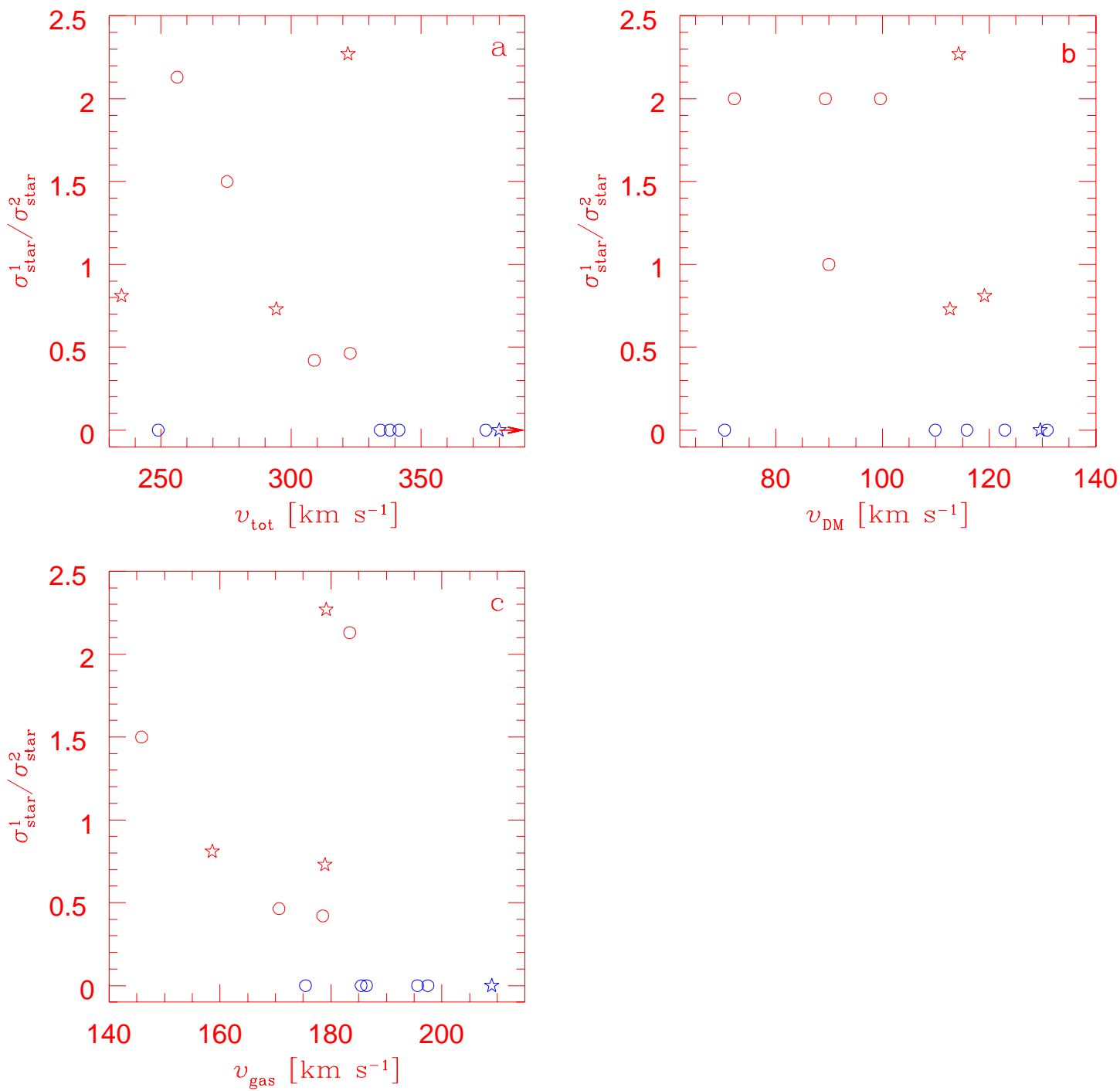}}
\caption{
 Ratio between the strengths of the
double starbursts ($\sigma_{\rm star}^{1}/\sigma_{\rm star}^{2}$)
versus the circular velocity at 2 kpc of the
a) total mass distribution ($\upsilon_{\rm tot}$),
b) dark matter ($\upsilon_{\rm DM}$) and c) the gas ($\upsilon_{\rm gas}$)
for all bursts in S.1 ({\it open circles\/}) and S.4 ({\it open stars\/}).
}
\end{figure}

\clearpage

\begin{figure}
{\psfig{file=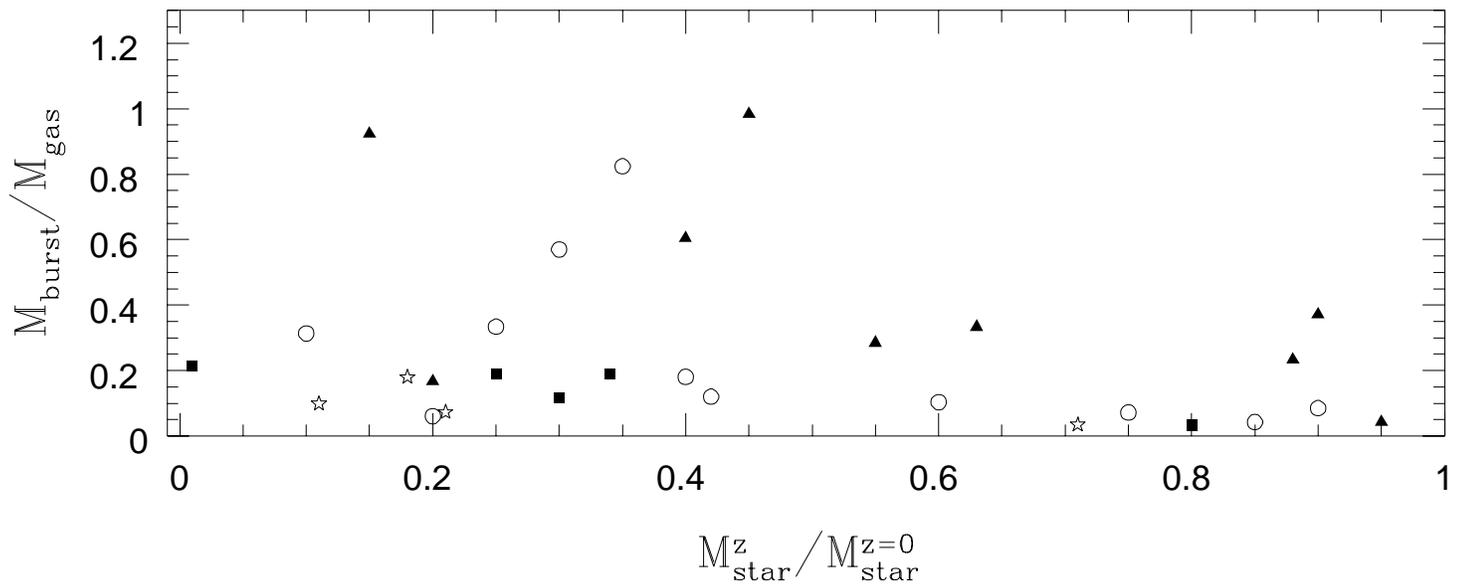}}
\caption{
The fraction of  gas consumed into stars during
starbursts, $M_{\rm burst}/M_{\rm gas}$, as a function of the fraction
of stars already in place in the progenitor 
at the $z$ of the merger ($M_{\rm star}^{z}/M_{\rm star}^{0}$).
Single bursts and primary components of double ones
have been included.
We find that, in a given simulation,  higher burst efficiencies
  tend to occur in  GLOs that lack an important stellar mass concentration
(see Fig.~3 for feature code).}
\end{figure}

\clearpage

\begin{figure}
{\psfig{file=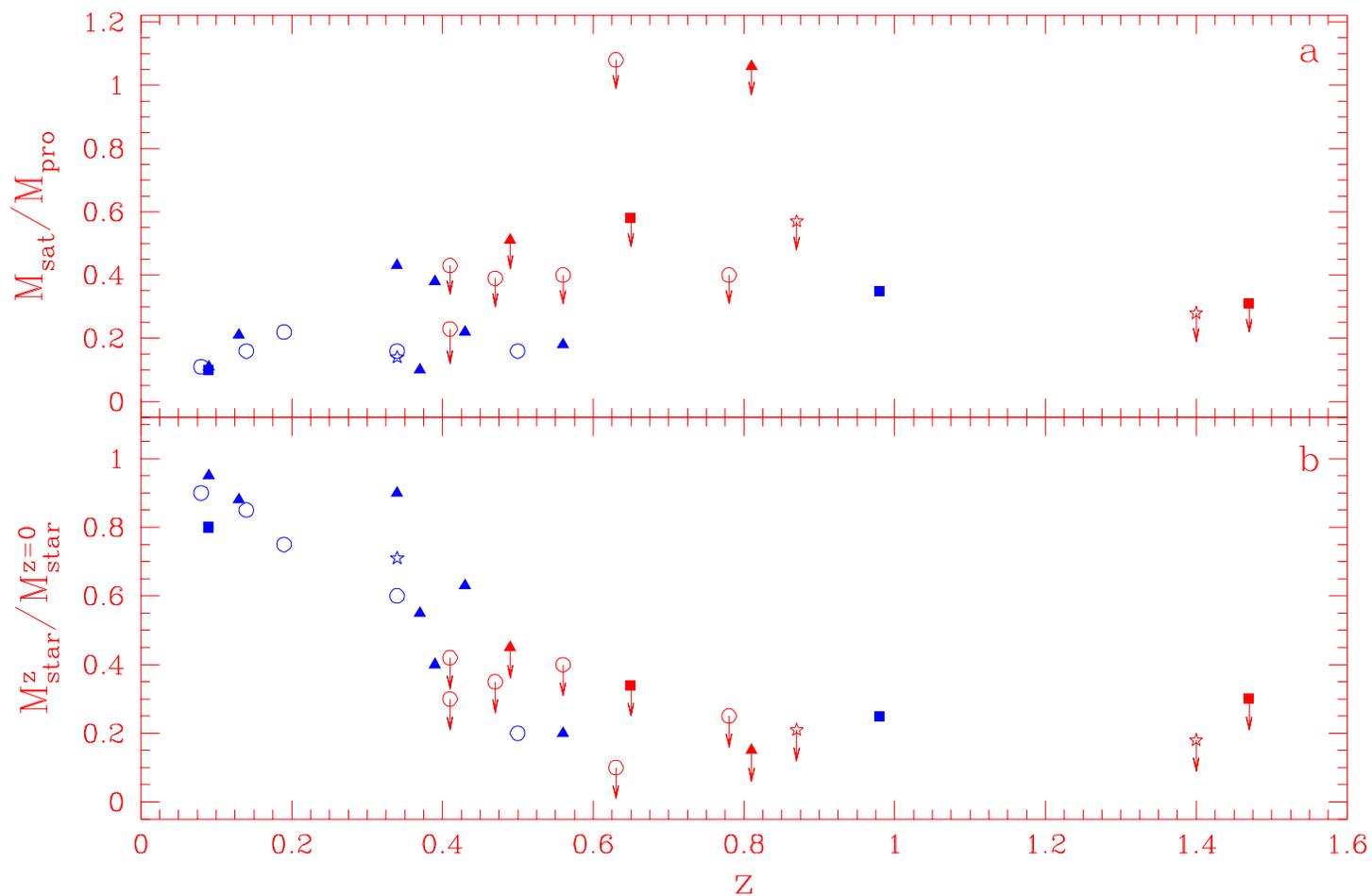}}
\caption{
 a) $M_{\rm sat}/M_{\rm pro}$ and b) $M_{\rm star}^{z}/M_{\rm star}^{z=0}$ as a function of the $z_{\rm burst}$ for single  and
primary components of double ones (symbols with arrows). See Fig.~3 for 
feature description. 
}
\end{figure}

\end{document}